\documentclass[fleqn,10pt]{wlscirep}
\title{Computational speed-up with a single qudit}

\author[1,*]{Z. Gedik}
\author[2]{I. A. Silva}
\author[1]{B. \c{C}akmak}
\author[3,4]{G. Karpat}
\author[2]{E. L. G. Vidoto}
\author[2]{D. O. Soares-Pinto}
\author[2]{E. R. deAzevedo}
\author[3]{F. F. Fanchini}

\affil[1]{Faculty of Engineering and Natural Sciences, Sabanci University, Tuzla, Istanbul, 34956, Turkey}
\affil[2]{Instituto de F\'isica de S\~ ao Carlos, Universidade de S\~ ao Paulo, Caixa Postal 369, 13560-970 S\~ ao Carlos, S\~ ao Paulo, Brazil}
\affil[3]{Faculdade de Ci\^encias, UNESP - Universidade Estadual Paulista, Bauru, S\~ ao Paulo, 17033-360, Brazil}
\affil[4]{Turku Center for Quantum Physics, Department of Physics and Astronomy, University of Turku, FIN-20014 Turku, Finland}

\affil[*]{gedik@sabanciuniv.edu}

\begin{abstract}
Quantum algorithms are known for providing more efficient solutions to certain computational tasks than any corresponding classical algorithm. Here we show that a single qudit is sufficient to implement an oracle based quantum algorithm, which can solve a black-box problem faster than any classical algorithm. For $2d$ permutation functions defined on a set of $d$ elements, deciding whether a given permutation is even or odd, requires evaluation of the function for at least two elements. We demonstrate that a quantum circuit with a single qudit can determine the parity of the permutation with only one evaluation of the function. Our algorithm provides an example for quantum computation without entanglement since it makes use of the pure state of a qudit. We also present an experimental realization of the proposed quantum algorithm with a quadrupolar nuclear magnetic resonance using a single four-level quantum system, i.e., a ququart.
\end{abstract}
\begin{document}

\flushbottom
\maketitle

\thispagestyle{empty}

\section*{Introduction}

Deutsch's algorithm was not only the first quantum algorithm but also one of the simplest \cite{Deutsch}. Although the algorithm was probabilistic in its original form, it has not been difficult to improve it to deterministic \cite{Cleve,Nielsen}. The Deutsch algorithm involves two qubits and distinguishes constant functions, which take both input values (0 or 1) to a single output value, from balanced functions in which output values are different. We introduce a simple algorithm that uses only a single qudit to determine the parity of chosen $2d$ permutations of a set of $d$ objects. As in the case of Deutsch's algorithm, we obtain a speedup relative to corresponding classical algorithms. For the particular computational task considered, the relative speedup starts from the case of a three-level quantum system, i.e., a qutrit.

What makes quantum algorithms interesting is that they can solve some problems faster than classical algorithms. Deutsch coined the term quantum parallelism to stress the ability of a quantum computer to perform two calculations simultaneously. How simple can a quantum circuit be? Or, what is the smallest quantum processor that can solve a problem faster than any classical algorithm? A closely related question is the origin of the power of quantum computation. Superposition, entanglement and discord are known to play essential roles in quantum computing and yet the origin of the power of the quantum algorithms is not completely clear \cite{vandernest}. Recently, it has been argued that quantum contextuality is a critical resource for quantum speedup of a fault tolerant quantum computation model \cite{Howard}. We present an example where an unentangled but contextual system can be used to solve a problem faster than classical methods. A qutrit is the smallest system where the contextual nature of quantum mechanics can be observed, in the sense that a particular outcome of a measurement cannot reveal the pre-existing definite value of some underlying hidden variable \cite{Kochen,Klyachko}. Whether the origin of the speedup of our algorithm can be explained by contextuality is an open question.

We present an oracle based quantum algorithm constructed on a surprisingly simple idea, which solves a black-box problem using only a single qudit without any correlation of quantum or classical nature. The black-box maps $d$ possible inputs to $d$ possible outputs after a permutation. The $2d$ possible permutation functions of $d$ objects are divided into two groups according to whether the permutation involves an odd or even number of exchange operations. The computational task is to determine the parity (oddness or evenness) of a given cyclic permutation. A classical algorithm requires two queries to the black-box. We show that a quantum algorithm can solve the problem with a single query. Even though the problem that the algorithm solves is not crucial, the algorithm is interesting in that it makes use of a single qudit, which means that neither entanglement nor any other correlation plays a role. Moreover, we present an experimental demonstration of this algorithm using a room temperature nuclear magnetic resonance (NMR) quadrupolar setup.

\section*{Results}

\subsection*{Computational task and the quantum algorithm}

Consider the case of three objects, where the six permutations of the set $\{1,2,3\}$ are (1,2,3), (2,3,1), (3,1,2), (3,2,1), (2,1,3), and (1,3,2). From the parity of the transpositions, the first three are even while last three are odd permutations. Our computational task is to determine the parity of a given permutation. If we treat a permutation as a function $f(x)$ defined on the set $x\in\{1,2,3\}$, determination of its parity requires evaluation of $f(x)$ for two different values of
$x$. We will show there exists a quantum algorithm where evaluating the function once (rather than twice) suffices to identify whether $f(x)$ is even or odd.

Since we are going to use standard spin operators in our discussion, let us denote the three states of a qutrit by $| m\rangle$, where $m=1,0,-1$ are the eigenvalue of $S_z$ with $S_z| m\rangle=m| m\rangle$. Rather than the permutations of the set $\{1,2,3\}$, we can then consider permutations of a possible $m$ values. Our aim here is to determine the parity of the bijection $f:\{1,0,-1\}\rightarrow\{1,0,-1\}$. We may define the three possible even functions $f_k$ using Cauchy's two-line notation
\begin{align}
f_1 =  \begin{pmatrix} 1&0&-1 \\ 1&0&-1 \\ \end{pmatrix},
f_2 =  \begin{pmatrix} 1&0&-1 \\ 0&-1&1 \\ \end{pmatrix},
f_3 =  \begin{pmatrix} 1&0&-1 \\ -1&1&0 \\ \end{pmatrix},
\end{align}
and the remaining three odd functions are
\begin{align}
f_4 =  \begin{pmatrix} 1&0&-1 \\ -1&0&1 \\ \end{pmatrix},
f_5 =  \begin{pmatrix} 1&0&-1 \\ 0&1&-1 \\ \end{pmatrix},
f_6 =  \begin{pmatrix} 1&0&-1 \\ 1&-1&0 \\ \end{pmatrix}.
\end{align}
Being a simple transposition of orthonormal states $| m\rangle$, the operator $U_{f_k}$ corresponding to $f_k$ is unitary and can be easily implemented. Direct application of $U_{f_k}$ on basis states does not bring any improvement on the classical solution, we still need to know the result of $U_{f_k}| m\rangle$ for two different values of $m$. However, quantum gates can act on any superposition state including the state $|\psi_1\rangle=\left(\exp[i2\pi/3]|1\rangle+|0\rangle+\exp[-i2\pi/3]|-1\rangle\right)/\sqrt{3}$. The state vector $|\psi_1\rangle$ can be obtained from $|1\rangle$ by the single qutrit Fourier transformation
\begin{equation}
U_{FT}= \frac{1}{\sqrt{3}}\begin{pmatrix}
\exp[i2\pi/3]&1&\exp[-i2\pi/3] \\
1&1&1 \\
\exp[-i2\pi/3]&1&\exp[i2\pi/3] \\
\end{pmatrix},
\end{equation}
in $S_z-$basis. We will show that this can be used to distinguish even and odd $f_k$'s. Note that the state vectors defined by $|\psi_k\rangle\equiv U_{f_k}|\psi_1\rangle=\left(\exp[i2\pi/3]|
f_k(1)\rangle+| f_k(0)\rangle+\exp[-i2\pi/3]| f_k(-1)\rangle\right)/\sqrt{3}$
have the property that $|\psi_1\rangle=\exp[-i2\pi/3]|\psi_2\rangle=\exp[i2\pi/3]|\psi_3\rangle$, and similarly $|\psi_4\rangle=\exp[-i2\pi/3]|\psi_5\rangle=\exp[i2\pi/3]|\psi_6\rangle$. Hence, application of $U_{f_k}$ on $|\psi_1\rangle=U_{FT}|1\rangle$ gives $|\psi_1\rangle$ for even $f_k$ and $|\psi_4\rangle=U_{FT}|-1\rangle$ for odd $f_k$. Therefore, if we apply the inverse Fourier transformation $U_{FT}^\dagger$ on $|\psi_k\rangle$, we have the state $|1\rangle$ (even $f_k$) or $|-1\rangle$ (odd $f_k$). Thus, a single evaluation of the function is enough to determine its parity.

\begin{figure}[t]
\centering
\includegraphics[scale=0.135]{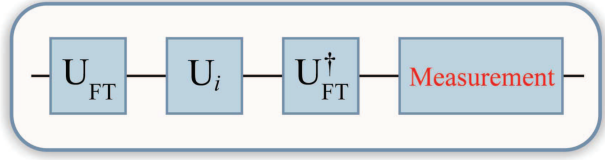}
\caption{Schematic view of the quantum circuit implementing the proposed quantum algorithm.}
\label{fig:fig1}
\end{figure}

In summary, the quantum circuit involves just three gates visited by a single qutrit. We start with $|1\rangle$ and place $U_{FT}$, $U_{f_k}$, and $U_{FT}^\dagger$ next to each other, as depicted in Fig. \ref{fig:fig1}. The final state of the qutrit after $U_{FT}^\dagger$ gate is necessarily either $|1\rangle$ or $|-1\rangle$, while $|0\rangle$ is never observed. Although we can modify our algorithm for a single qubit, where the Fourier transformation becomes a Hadamard operator, this case is not interesting since the classical solution requires only a single evaluation of the permutation function so the quantum algorithm does not provide any speedup. The qutrit case of our algorithm is one of the simplest quantum algorithms.

\begin{figure*}[t]
\centering
\includegraphics[scale=0.2]{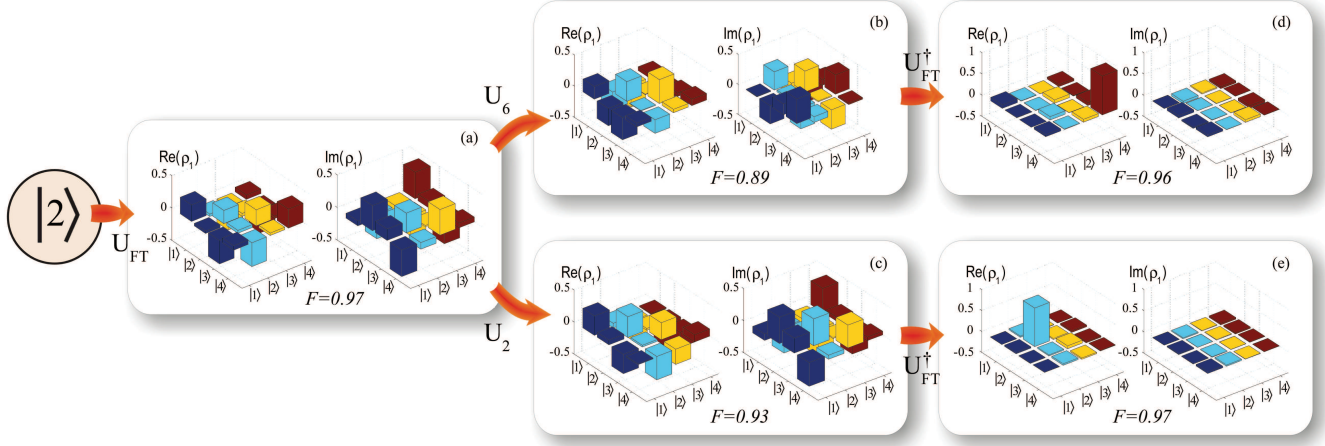}
\caption{Experimental demonstration of the algorithm. We create the initial state $|2\rangle$ with a fidelity of $0.99$. From left to right is a bar representation of the density matrix for the state after the application of the Fourier transformation, $U_{FT}$.  (a) Obtained by quantum state tomography. (b) Applying the pulses that implement $U_{6}$. (c) Applying the pulses that implement  $U_{2}$ (c). The two possible outcomes of the algorithm (d) $|4\rangle$ for negative and $|2\rangle$ for positive cyclic permutations. The experimental errors were quantified by the relation between signal and signal-to-noise ratio. For all of the reconstructed density matrices, the errors are always smaller than $6\%$ (see Supplementary Material for details).}
\label{fig:fig2}
\end{figure*}

We can generalize the algorithm to $d$ dimensional (or equivalently spin $(d-1)/2$) systems. In that case, the algorithm may be used to distinguish cyclic permutations according to their parity. For example, when $d=4$ positive cyclic permutations of $(1,2,3,4)$ are $(2,3,4,1)$, $(3,4,1,2)$ and $(4,1,2,3)$ while the negative cyclic permutations are $(4,3,2,1)$, $(3,2,1,4)$, $(2,1,4,3)$ and $(1,4,3,2)$. As in the case of three elements, given one of the eight permutations, our aim is to determine its parity and this requires knowing at least two elements in the permutation, or, equivalently, knowing the values of the function for two variables classically.

For a four level quantum system (ququart), we can use the initial state $|\psi_2\rangle=\left(|1\rangle+i|2\rangle-|3\rangle-i|4\rangle\right)/2$ where $|k\rangle$'s are states of the ququart with vector representations $|1\rangle=(1, 0, 0, 0)^T$, $|2\rangle=(0, 1, 0, 0)^T$, $|3\rangle=(0, 0, 1, 0)^T$, and $|4\rangle=(0, 0, 0, 1)^T$. In this case we can use the standard quantum Fourier transformation \cite{Nielsen} which can be viewed as a unitary matrix
\begin{equation}
U_{FT}=\frac{1}{2}
\begin{pmatrix}
1&1&1&1\\
1&i&-1&-i\\
1&-1&1&-1\\
1&-i&-1&i\\
\end{pmatrix}
\end{equation}
in $|k\rangle-$basis, so that $|\psi_2\rangle=U_{FT}|2\rangle$. Observe that, starting from $(1,2,3,4)$,  the positive cyclic permutations $(1,2,3,4)$, $(2,3,4,1)$, $(3,4,1,2)$, and $(4,1,2,3)$ can be obtained with the corresponding unitary matrices,
\begin{align}
 U_1 &= \begin{pmatrix} 1 & 0 & 0 & 0 \\ 0 & 1 & 0 & 0 \\ 0 & 0 & 1 & 0 \\ 0 & 0 & 0 & 1 \end{pmatrix}, &
 U_2 &= \begin{pmatrix} 0 & 0 & 0 & 1 \\ 1 & 0 & 0 & 0 \\ 0 & 1 & 0 & 0 \\ 0 & 0 & 1 & 0 \end{pmatrix}, & \nonumber \\
 U_3 &= \begin{pmatrix} 0 & 0 & 1 & 0 \\ 0 & 0 & 0 & 1 \\ 1 & 0 & 0 & 0 \\ 0 & 1 & 0 & 0 \end{pmatrix}, &
 U_4 &= \begin{pmatrix} 0 & 1 & 0 & 0 \\ 0 & 0 & 1 & 0 \\ 0 & 0 & 0 & 1 \\ 1 & 0 & 0 & 0 \end{pmatrix}, &
\end{align}
respectively, and they map $|\psi_2\rangle$ onto $|\psi_2\rangle$, $-i|\psi_2\rangle$, $-|\psi_2\rangle$, and $i|\psi_2\rangle$.  On the other hand, the negative cyclic permutations result in $-i|\psi_4\rangle$, $-|\psi_4\rangle$, $i|\psi_4\rangle$, and $|\psi_4\rangle$,
which can be similarly realized by the unitary matrices,
\begin{align}
 U_5 &= \begin{pmatrix} 0 & 0 & 0 & 1 \\ 0 & 0 & 1 & 0 \\ 0 & 1 & 0 & 0 \\ 1 & 0 & 0 & 0 \end{pmatrix}, &
 U_6 &= \begin{pmatrix} 0 & 0 & 1 & 0 \\ 0 & 1 & 0 & 0 \\ 1 & 0 & 0 & 0 \\ 0 & 0 & 0 & 1 \end{pmatrix}, & \nonumber \\
 U_7 &= \begin{pmatrix} 0 & 1 & 0 & 0 \\ 1 & 0 & 0 & 0 \\ 0 & 0 & 0 & 1 \\ 0 & 0 & 1 & 0 \end{pmatrix}, &
 U_8 &= \begin{pmatrix} 1 & 0 & 0 & 0 \\ 0 & 0 & 0 & 1 \\ 0 & 0 & 1 & 0 \\ 0 & 1 & 0 & 0 \end{pmatrix}. &
\end{align}
respectively, where $|\psi_4\rangle=U_{FT}|4\rangle$. Therefore, applying the inverse Fourier transformation $U_{FT}^\dagger$ and checking the final state of the ququart, we can determine the parity of the cyclic permutation. Final state $|2\rangle$ indicates that the permutation is even, while $|4\rangle$ means that the permutation is odd. As in the case of qutrit, the quantum algorithm allows us to determine the parity of a cyclic permutation with a single evaluation of the permutation function rather than the two that would be required classically. For four elements, we can formulate two more examples using other circular permutations, and evaluate these new cases by redefining the Fourier transformation. For example, the positive $(1,3,2,4)$, $(3,2,4,1)$, $(2,4,1,3)$, $(4,1,3,2)$, and negative $(4,2,3,1)$, $(2,3,1,4)$, $(3,1,4,2)$ cyclic permutations can be distinguished with a single evaluation provided we start with the state $|\psi_2\rangle=\left(|1\rangle-|2\rangle+i|3\rangle-i|4\rangle\right)/2$. The last eight members of the total $4!=24$ permutations can also be used to set up a similar problem. Moving to a $d-$level quantum system (qudit), we can define
\begin{equation}
|\psi_k\rangle=\frac{1}{\sqrt{d}}\sum_{k'=1}^{d}\exp\left[i\frac{2\pi}{d}(k-1)(k'-1)\right]
| k'\rangle.
\end{equation}
In this case, the positive cyclic permutations map $|\psi_2\rangle$ onto itself while the negative permutations give $|\psi_d\rangle$.

From the above generalizations, we deduce that the essence of the algorithm is to design a circuit so that output states are grouped according to the computational task where final states are described by the same vectors up to a phase factor. For this type of generalization, the speedup factor will be two as in the case of a single qutrit. We can look for further generalizations of the algorithm with larger or perhaps exponential speedup factors by using many qudits together. However, the main purpose of the present work is to find the simplest quantum system which provides a computational speedup. Thus, we can identify the minimum system requirements for a useful quantum algorithm.

\subsubsection*{Experimental demonstration}

In the following, we present an experiment which demonstrates the quantum algorithm for a ququart. Historically, many quantum algorithms were implemented in NMR systems \cite{cory1997,cory2000,marx2000,cory2004,suter2008,jones2011}, especially those algorithms where entanglement is not required \cite{du2001,dqc1,laflamme2002,passante2011}. The implementation of the algorithm using a ququart is achieved using a spin--$\frac{3}{2}$ nuclei, which has been extensively used in NMR-QIP applications as exemplified in \cite{fung2000,kumar2001,fung2002,kumar2003,dudu2004,dudu2005,murali2004,kampermann2005,gopinath2008,kondo2008,tan2012,dosp2012,araujo2013} and reviewed in \cite{teles2012}. In such NMR systems, a strong static magnetic field is responsible for the Zeeman splitting, providing four energy levels. Since the nuclear spin is $I > \frac{1}{2}$, the nuclei possess a quadrupole moment that interacts with the electric field gradient created by the surrounding charge distribution, i.e., quadrupolar interaction.  When this interaction is much stronger than the quadrupolar one, we can use perturbation theory and express the Hamiltonian as \cite{abra}
\begin{align}
H = -\hbar\omega_{L} I_{z} + \frac{\hbar\omega_{Q}}{6}\,(3\,I_{z}^{2} - \bold{I}^{2}),
\end{align}
where $\omega_{L}$ is the Larmor frequency, $\omega_{Q}$ is the quadrupolar frequency ($|\omega_L| \gg |\omega_Q|$), $I_{z}$ is the $z$ component of the nuclear spin operator, and $\bold{I}$ is the total nuclear spin operator. The eigenstates of the system are by $\left|3/2\right\rangle$, $\left|1/2\right\rangle$, $\left|-1/2\right\rangle$, and $\left|-3/2\right\rangle$,  indexed as $|1\rangle$, $|2\rangle$, $|3\rangle$, $|4\rangle$, respectively. The corresponding NMR spectrum is composed by three spectral lines associated to the three single quantum transitions, $\Delta m \pm 1$.

The initial state is prepared from the thermal equilibrium state using a time averaging procedure based on numerically optimized radio frequency (rf) pulses generally called strong modulating pulses (SMP) \cite{fortunato,kampermann2005,teles2007}. The technique consists of using blocks of concatenated rf pulses, with amplitudes, phases, and durations optimized to provide a state preparation such that density matrix is $\rho=\frac{(1-\epsilon)}{4}\,\mathbb{I}_{4}+\epsilon\rho_1=\frac{(1-\epsilon)}{4}\,\mathbb{I}_{4}+\epsilon|i\rangle\langle i|$, where $\rho_1$ is a trace one density matrix corresponding to the state $|i\rangle\langle i|$ defined by the optimized SMP pulses \cite{ivan}. The quantum gates in the circuit are also implemented using these SMP optimized pulses. Since NMR measurements are not sensitive to the identity part of the density matrix, the term $|i\rangle\langle i|$ is manipulated and read out selectively. The SMP optimization technique is based on the Nelder-Mead Simplex minimization method which is explained in detail in \cite{Nelder}.

The steps of the protocol were implemented as follows: ($i$) we apply the SMP optimized gate $U_{FT}$ to the initial state $|2\rangle$ to obtain $|\psi_2\rangle$; ($ii$) we apply the SMP optimized gate $U_iU_{FT}$ for $i = 2, 6$ to the initial state again; ($iii$) finally, starting once more from the initial state, we implement the SMP optimized gate $U^{\dagger}_{FT}U_iU_{FT} $ for $i = 2, 6$ to obtain either $|2\rangle$ or $|4\rangle$ as an outcome of the algorithm, as schematically depicted in Fig. \ref{fig:fig1}. Figure \ref{fig:fig2} shows a bar representation of the density matrices, after each step of the protocol, obtained by quantum state tomography.

\section*{Discussion}

We have shown that a single qudit can be used to implement a quantum algorithm which provides a two to one speedup in determining parity of cyclic permutations. Even though the model problem is not one of the most important computational tasks and the speedup is not exponential when generalized to higher dimensional cases, the algorithm is still important since it provides a strikingly simple example for quantum computation without entanglement.

We have experimentally demonstrated the proposed algorithm using a quadrupolar NMR setup, and showed that it deterministically decides whether a given permutation, from a set of eight possible functions, of four objects is positive or negative cyclic with a single query to the black-box.

Despite the simplicity of the algorithm, the origin of the speedup remains unclear. It is evident that quantum correlations do not supply the solution of the computational task since a single quantum system is considered. Regardless, the true resource behind the power of this algorithm remains an open question.

\section*{Methods}

For our experimental system, as for all room temperature NMR, the density matrix can be expressed as  $\rho=\frac{1}{4}\mathbb{I}_{4}+\epsilon\Delta\rho$, where $\epsilon=\hbar\omega_L /4k_BT \sim 10^{-5}$ is the ratio between the magnetic and thermal energies, $\omega_L$ is the Larmor frequency, $k_B$ is the Boltzmann constant, and $T$ the temperature \cite{abra,ivan}. Measurements and unitary transformations only affect the traceless deviation matrix, $\Delta\rho$, which contains all the available information about the state of the system. Unitary transformations over $\Delta\rho$ are implemented by radio frequency pulses and/or evolutions under spin interactions, with excellent control of rotation angle and direction. The full characterization of $\Delta\rho$ can be achieved using many available quantum state tomography protocols \cite{long2001,kampermann2002,leskowitz2004,teles2007}. Since for NMR experiments only the deviation matrix is detected, density matrix elements are expressed in units of $\epsilon$.

The experiment was performed using sodium nuclei, $^{23}$Na, in a lyotropic liquid crystal sample at room temperature. The sample was prepared with 20.9 wt\% of Sodium Dodecyl Sulfate (SDS) (95\% of purity), 3.7 wt\% of decanol, and 75.4 wt\% of deuterium oxide, following the procedure in\cite{amostra}. The $^{23}$Na NMR experiments were performed in a 9.4-T VARIAN INOVA spectrometer using a 5 mm solid state NMR probe head at $T = 25^{º}$C. We obtained the quadrupole frequency $\nu_{Q} = \omega_{Q} / 2 \pi = 10$ kHz. Under the conditions of the experiment, the sample can be considered as an ensemble of isolated sodium nuclei, i.e., an ensemble of individual ququarts.

The reconstruction of density matrices was performed  using the method described in \cite{teles2007}, based on a coherence selection procedure, i.e., read out pulses with specifically designed amplitudes, durations, and phases were applied to obtain an NMR spectrum associated only with the density matrix elements of a specific coherence order. The three line intensities of this spectrum ($I_i$) were used as inputs to a set of equations, which provided the selected density matrix elements. To estimate the experimental uncertainties, we assumed the error associated to each spectral line ($\Delta I$) to be the standard deviation of the spectral noise obtained from the signal-to-noise ratio.  Hence, the maximum and minimum threshold of each line was calculated as $I_{i,max} = I_i  + \Delta I $ and $I_{i,min} = I_i  - \Delta I $. We reconstructed the density matrix considering all possible combinations of maximum and minimum intensities, resulting in a set of reconstructed density matrices, and the mean and standard deviation of each density matrix element were obtained. The elements of the average density matrices and their respective errors are shown in the Supplementary Material. All relative errors are smaller than $6\%$. The fidelities to the theoretical predictions are shown in Fig. \ref{fig:fig2}.

After the completion of this work, we became aware of subsequent works, also implementing the quantum algorithm proposed here but in different experimental setups \cite{dogra,zhan,wang}.

\section*{Acknowledgements}

ZG is supported by the Turkish agency TUBITAK under Grant No. 111T232. BC is supported by Sao Paulo Research Foundation (FAPESP) under grant No. 2014/21792-5. GK is supported by S\~{a}o Paulo Research Foundation (FAPESP), BEPE fellowship, given under grant no 2014/20941-7. DOSP is supported by the National Counsel of Technological and Scientific Development (CNPq) under grants numbers 304955/2013-2 and 443828/2014-8, and by Coordena\c{c}\~{a}o de Aperfei\c{c}oamento de Pessoal de N\'{i}vel Superior (CAPES) under the grant number 108/2012. ERdeA is supported by the National Counsel of Technological and Scientific Development (CNPq) under grant number 312852/2014-2. FFF is supported by S\~{a}o Paulo Research Foundation (FAPESP) under grant number 2012/50464-0, and by the National Counsel of Technological and Scientific Development (CNPq) under grant number 474592/2013-8. DOSP and FFF are members of the Brazilian National Institute of Science and Technology of Quantum Information (INCT/IQ). The authors would like to thank A. Levi, C. Sa\c{c}l{\i}o\u{g}lu, J. Wallman, and J. Wrachtrup for helpful discussions.

\section*{Author contributions statement}

Z.G. developed the theoretical ideas and proposed the presented quantum algorithm, I.A.S, E.L.G.V., D.O.S.P. and E.R.deA. conducted the experiment and analyzed the data, B.\c{C}., G.K. and F.F.F. coordinated the project, Z.G., B.\c{C}., G.K., I.A.S, D.O.S.P. and E.R.deA. wrote the manuscript.  All authors reviewed the manuscript.

\section*{Additional information}

\textbf{Competing financial interests:} The authors declare no competing financial interests.


\begin{thebibliography}{99}

\bibitem{Deutsch} Deutsch, D. Quantum theory, the Church-Turing principle and the universal quantum computer. \textit{Proc. R. Soc. Lond. A} \textbf{400}, 97 (1985).
\bibitem{Cleve} Cleve, R., Ekert, A., Macchiavello, C. \& Mosca, M. Quantum algorithms revisited. \textit{Proc. R. Soc. Lond. A} \textbf{454}, 339 (1998).
\bibitem{Nielsen} Nielsen, M. A. \& Chuang, I. L. \textit{Quantum Computation and Quantum Information} (Cambridge University Press, Cambridge, England, 2000).
\bibitem{vandernest} van den Nest, M. Universal quantum computation with little entanglement. \textit{Phys. Rev. Lett.} \textbf{110}, 060504 (2013).
\bibitem{Howard} Howard, M., Wallman, J., Veitch, V. \& Emerson J. Contextuality supplies the 'magic' for quantum computation. \textit{Nature} \textbf{510}, 351 (2014).
\bibitem{Kochen} Kochen, S. \& Specken, E. P. The problem of hidden variables in quantum mechanics. \textit{J. Math. Mech.} \textbf{17}, 59 (1967).
\bibitem{Klyachko} Klyachko, A. A., Can, M. A., Binicio\u{g}lu, S. \& Shumovsky, A. S. Simple test for hidden variables in spin-1 systems. \textit{Phys. Rev. Lett.} \textbf{101}, 020403 (2008).
\bibitem{cory1997} Cory, D. G., Fahmy, A. F. \& Havel, T. F. Ensemble quantum computing by NMR spectroscopy. \textit{Proceedings of the National Academy of Sciences} USA \textbf{94}, 1634 (1997).
\bibitem{cory2000} Cory, D. G. et al.  NMR based quantum information processing: Achievements and prospects. \textit{Fortschritte der Physik} \textbf{48}, 875 (2000).
\bibitem{marx2000} Marx, R., Fahmy, A. F., Myers, J. M., Bermel, W. \& Glaser, S. J. Approaching five-bit NMR quantum computing. \textit{Phys. Rev. A} \textbf{62}, 012310 (2000).
\bibitem{cory2004} Ramanathan, C. et al. NMR quantum information processing. \textit{Quantum Inf. Process} \textbf{3}, 15 (2004).
\bibitem{suter2008} Suter, D. \& Mahesh, T. S. Spins as qubits: quantum information processing by nuclear magnetic resonance. \textit{J. Chem. Phys.} \textbf{128}, 052206 (2008).
\bibitem{jones2011} Jones, J. A. Quantum Computing with NMR. \textit{Prog. NMR Spectrosc.} \textbf{59}, 91 (2011).
\bibitem{du2001} Du, J. et al. Implementation of a quantum algorithm to solve the Bernstein-Vazirani parity problem without entanglement on an ensemble quantum computer. \textit{Phys. Rev. A} \textbf{64}, 042306 (2001).
\bibitem{dqc1} Knill, E. \& Laflamme, R. Power of One Bit of Quantum Information. \textit{Phys. Rev. Lett.} \textbf{81}, 5672 (1998).
\bibitem{laflamme2002} Laflamme, R., Cory, D. G., Negrevergne, C. \& Viola, L. NMR quantum information processing and entanglement. \textit{Quant. Inf. Comp.} \textbf{2}, 166 (2002).
\bibitem{passante2011} Passante, G., Moussa, O., Trottier, D. A. \& Laflamme, R. Experimental detection of nonclassical correlations in mixed-state quantum computation. \textit{Phys. Rev. A} \textbf{84}, 044302 (2001).
\bibitem{fung2000} Khitrin, A. K. \& Fung, B. M. Nuclear magnetic resonance quantum logic gates using quadrupolar nuclei. \textit{J. Chem. Phys.} \textbf{112}, 6963 (2000).
\bibitem{kumar2001} Sinha, N., Mahesh, T. S., Ramanathan, K. V. \& Kumar, A. Toward quantum information processing by nuclear magnetic resonance: Pseudopure states and logical operations using selective pulses on an oriented spin 3/2 nucleus. \textit{J. Chem. Phys.} \textbf{114}, 4415 (2001).
\bibitem{fung2002} Ermakov, V. L. \& Fung, B. M. Experimental realization of a continuous version of the Grover algorithm. \textit{Phys. Rev. A}  \textbf{66}, 042310 (2002).
\bibitem{kumar2003} Das, R. \& Kumar, A. Use of quadrupolar nuclei for quantum-information processing by nuclear magnetic resonance: Implementation of a quantum algorithm. \textit{Phys. Rev. A}  \textbf{68}, 032304 (2003).
\bibitem{dudu2004} Bonk, F. A. et al. Quantum-state tomography for quadrupole nuclei and its application on a two-qubit system. \textit{Phys. Rev. A} \textbf{69}, 042322 (2004).
\bibitem{dudu2005} Bonk, F. A. et al.Quantum logical operations for spin 3/2 quadrupolar nuclei monitored by quantum state tomography. \textit{J. Mag. Res.} \textbf{175}, 226 (2005).
\bibitem{murali2004} Murali, K. V. R. M., Son, H. -B., Steffen, M., Judeinstein, P. \& Chuang, I. L. Test by NMR of the phase coherence of electromagnetically induced transparency. \textit{Phys. Rev. Lett.} \textbf{93}, 033601 (2004).
\bibitem{kampermann2005} Kampermann, H. \& Veeman, W. S. Characterization of quantum algorithms by quantum process tomography using quadrupolar spins in solid-state nuclear magnetic resonance. \textit{J. Chem. Phys.}  \textbf{122}, 214108 (2005).
\bibitem{gopinath2008} Gopinath, T. \& Kumar, A. Implementation of controlled phase shift gates and Collins version of Deutsch–Jozsa algorithm on a quadrupolar spin-7/2 nucleus using non-adiabatic geometric phases. \textit{J. Magn. Reson.}  \textbf{193}, 168 (2008).
\bibitem{kondo2008} Kondo, Y. et al. Multipulse operation and optical detection of nuclear spin coherence in a GaAs/AlGaAs quantum well. \textit{Phys. Rev. Lett.}  \textbf{101}, 207601 (2008).
\bibitem{tan2012} Tan, Y. P. et al. Preparing Pseudo-Pure States in a Quadrupolar Spin System Using Optimal Control. \textit{Chinese Phys. Lett.} \textbf{29}, 127601 (2012).
\bibitem{dosp2012} Araujo-Ferreira, A. G. et al. Quantum state tomography and quantum logical operations in a three qubits NMR quadrupolar system. \textit{Int. J. Quantum Inf.} \textbf{10}, 1250016 (2012).
\bibitem{araujo2013} Araujo-Ferreira, A. G. et al. Classical bifurcation in a quadrupolar NMR system. \textit{Phys. Rev. A} \textbf{87}, 53605, (2013).
\bibitem{teles2012} Teles, J. et al. Quantum information processing by nuclear magnetic resonance on quadrupolar nuclei. \textit{Philos. T. R. Soc. A} \textbf{370}, 4770, (2012).
\bibitem{abra} Abragam, A. {\it The Principles of Nuclear Magnetism}, (Oxford University Press, 1978).
\bibitem{fortunato} Fortunato, E. M. et al. Design of strongly modulating pulses to implement precise effective Hamiltonians for quantum information processing. \textit{J. Chem. Phys.} \textbf{116}, 7599 (2002).
\bibitem{teles2007} Teles, J. et al. Quantum state tomography for quadrupolar nuclei using global rotations of the spin system. \textit{J. Chem. Phys.} \textbf{126}, 154506 (2007).
\bibitem{ivan} Oliveira, I. S., Bonagamba, T. J., Sarthour, R. S., Freitas, J. C. C. \& deAzevedo, E. R. {\it NMR Quantum Information Processing} (Elsevier, Amsterdam, 2007).
\bibitem{Nelder} Nelder, J. A. \& Mead, R. A simplex method for function minimization. \textit{Computer Journal} \textbf{7}, 308 (1965).
\bibitem{long2001} Long, G. L., Yan, H. Y. \& Sun, Y. Analysis of density matrix reconstruction in NMR quantum computing. \textit{J. Opt. B} \textbf{3}, 376 (2001).
\bibitem{kampermann2002} Kampermann, H. \& Veeman, W. S. Quantum computing using quadrupolar spins in solid state NMR. \textit{Quantum Inf. Process} \textbf{1}, 327 (2002).
\bibitem{leskowitz2004} Leskowitz, G. M. \& Mueller, L. J. State interrogation in nuclear magnetic resonance quantum-information processing. \textit{Phys. Rev. A} \textbf{69}, 052302 (2004).
\bibitem{amostra} Radley, K., Reeves, L. W. \& Tracey, A. S. Effect of counterion substitution on the type and nature of nematic lyotropic phases from nuclear magnetic resonance studies. \textit{J. Phys. Chem.} \textbf{80}, 174 (1976).
\bibitem{dogra} Dogra S., Arvind, \& Dorai K. Determining the parity of a permutation using an experimental NMR qutrit, \textit{Phys. Lett. A} \textbf{378}, 3452 (2014).
\bibitem{zhan} Zhan X., Li J., Qin H., Bian Z. \& Xue P.  Linear optical demonstration of quantum speed-up with a single qudit, \textit{Opt. Express} \textbf{23} 18422 (2015).
\bibitem{wang} Wang F. et al. Demonstration of quantum permutation algorithm with a single photon ququart, \textit{Sci. Rep.} \textbf{5} 10995 (2015).


\end{thebibliography}
\end{document}